\documentclass[preprint,doublespace,showpacs]{revtex4} 

\usepackage{graphicx}

\usepackage{dcolumn}
\usepackage{bm}
\usepackage{amssymb}
\usepackage{amsmath}

\begin{document}

\title{All-optical periodic code matching by a single-shot frequency-domain cross-correlation measurement}

\author{Lev Chuntonov, Leonid Rybak, Andrey Gandman,}
\author{Zohar Amitay}
\email{amitayz@tx.technion.ac.il} %
\affiliation{Schulich Faculty of Chemistry, Technion - Israel Institute of Technology, Haifa 32000, Israel}

\begin{abstract}
Optical single-short measurement of the cross-correlation function
between periodic sequences is demonstrated.
The sequences are encoded into the broadband ultrashort phase-shaped
pulses which are mixed in the nonlinear medium with additional
amplitude-shaped narrowband pulse.
The spectrum of the resulted four wave mixing signal is measured to
provide the cross-correlation function.
The high contrast between the values of cross-correlation and
auto-correlation (the latter includes also the information of the
sequence period) has potential to be employed in the optical
implementation of CDMA communication protocol.
\end{abstract}

\maketitle

Modern communication technology, including the mobile 3G wide-band
and Global Positioning System standards, implemented in the
rf-regime, extensively employs communication based on the Code
Division Multiple Access (CDMA) protocol \cite{cdma_book}.
It is known to be preferable over the Time Division and Frequency
Division multiplexing strategies, where different time or frequency
slots, respectively, are assigned to different users sharing the
common communication channel.
In CDMA the transmitter and receiver are assigned a unique code to
recognize each other from the overall information flow.
The codes are converted into the individual characteristics of
different frequency components of the broadband spectrum and
translated to the time-dependence of the waveform.
This highly efficient spread spectrum multiplexing strategy is
potentially capable to support a high number of users, given a large
codes families with low correlation properties are available
\cite{coding_theory_handbook}.
These code families, thus, must support sufficient number of users,
while minimizing the interference between them.
The state of the art code families \cite{kumar_1, kumar_2, kumar_3}
incorporate periodic cyclically distinct pseudo-random sequences
over the arbitrary alphabet.
The codes have low correlating properties which imply minimal
interference between them.
This generalization to the extended alphabet dramatically improved
the capacity and performance of the CDMA networks, while in the past
only the binary codes were considered.

It was recognized several decades ago that the developments of the
communication technology in the rf region are also applicable for
the optical communication \cite{ocdma_handbook, ocdma_weiner}.
In the optical implementation the codes are spread over the broad
bandwidth of the light pulses, which are phase-shaped in the
frequency-domain \cite{pulse_shaping-weiner}.
Recently, the demand for the fast and reliable multiuser
communication stimulated an extensive research activity in this
field, focusing on different aspects of optical networking
\cite{ocdma_handbook, ocdma_weiner}.
Availability of a powerful technique to discriminate between
different codes is critical to fully benefit from the advantages of
CDMA \cite{cross-corr-weiner-1, cross-corr-weiner-2}.
Here we demonstrate efficient all-optical experimental recognition
of matching codes from the large family using single-shot
measurement of their entire cross-correlation function in the
frequency-domain.
The interference between the photo-induced optical pathways excited
with two encoded broadband pulses and one amplitude-shaped
narrowband pulse within the non-linear medium is employed to
evaluate the cross-correlation function between the codes.
We utilize the BOXCARS setup \cite{boxcars_eckbreth} to obtain
spatially-resolved background-free four-wave-mixing spectrum which
is formally equivalent to the cross-correlation function between the
codes of interest.
Our approach enables the high-contrast discrimination between the
periodic low-correlating codes based on identification of the
constructive interference corresponding to the pulses shaped with
similar codes and destructive interference corresponding to the
pulses shaped with different codes.
As opposed to the previously developed techniques
\cite{cross-corr-weiner-1,cross-corr-weiner-2}, the presented setup
enables single-shot measurement of the entire cross-correlation
function, which in addition provides usefull information on the
unknown period of the code without additional computational
expences.

We consider the families of periodic cyclically distinct
pseudo-random sequences $s_{k}(t)$ over the alphabet of integers of
size $p$ $\left[0\le s_k(t) \le p-1 \right]$ with arithmetics modulo
$p$, where $p\ge$2.
The cross-correlation function between two sequences is defined by
\begin{equation}\label{eq1}
c(k,r,\tau)=\sum_{t=0}^{L-1}\omega_{p}^{s_{k}(t+\tau)-s_{r}(t)},
\end{equation}
where $k$ and $r$ are indexes of the sequences in the set of size
$M$, $0\le\tau\le L-1$, $L$ - is the period of the sequence, and
$\omega_{p}=e^{i2\pi/p}$ - is $p$th root of unity.
The commonly used merit factor of the codes performance is the
maximal value of the correlation $C_{max}$=max$\left\{
c(k,r,\tau)\right\}$ excluding the cases where simultaneously $k=r$
and $\tau$=0, for which, trivially, $C_{max}=L$.
We thus demand $C_{max}\ll L$ to ensure the non-intercepted
communication with the absence of interference.
The nested chain low-correlation codes families $S(0)\subseteq S(1)
\subseteq S(2)\subseteq \ldots$ obey the desired requirements
\cite{kumar_1, kumar_2, kumar_3}.
For a given period $L$, the code family of higher order incorporate
more code sequences, while the corresponding value of $C_{max}$
increases.
However, as the code period $L$ increases, the number of available
codes increases, while the value of $C_{max}$ decreases.
The $C_{max}$ values of the $S$-families have a well defined bounds
and distribution, and thus can be appropriately tailored to the
implementation needs by the choice of $L$ and the order of the
family.

In the optical domain, the common implementation of the CDMA network
communication involves broadband ultra-fast shaped laser pulses .
The high popularity of this benchmark system is mainly due to the
Weiner-Heritage design of the pulse shaping apparatus that is used
to assign the codes to the laser pulses carrying the transmitted
data.
The 4f optical setup of the pulse shaper incorporates pixelated
liquid crystal Spatial Light Modulator (SLM) in its focal plane,
where each spectral component of the laser electric field
$E(\omega)=\left|E(\omega)\right|e^{i\Phi(\omega)}$ is addressed
individually.
Here, $\left|E(\omega)\right|$ - is the spectral amplitude at
frequency $\omega$, and $\Phi(\omega)$ - is the corresponding phase.
The phase-shaping is implemented via control over the voltage
applied to different pixels of liquid crystal, which induces
different phase retardation of the transmitted light.
We assume that spectral segment of $\Delta\omega$ is spread along
each pixel.
The similarity between the representation of integer code sequences
$s_{k}(\tau)$ in the complex plane and the phase factor of the
electric field is used to convert the codes into the relative phases
of the broadband pulse spectral components:
$\Phi(\omega)=(2\pi/p)s_{k}(\omega)$.
Once the encoding process is completed at the transmitter site,
measurement of the cross-correlation function $c(k,r,\tau)$ is the
important step that should be performed at the receiver or router
site.

Naturally, the single-shot all-optical cross-correlation measurement
is the ideal way to effectively discriminate between different codes
and identify their period.
Translation of the codes into the phases of the pulse spectral
components allows one to take advantage of the interference between
multiple photo-induced multi-photon pathways excited in the
non-linear medium by the broadband pulse.
The cross-correlation function $c(k,r,\tau)$ between the two codes
$s_{k}$ and $s_{r}$ converted into the phases $\Phi_{k}(\omega)$ and
$\Phi_{r}(\omega)$ of the pulses with the electric fields
$E_{k}(\omega)$ and $E_{r}(\omega)$ obtains the form of
\begin{equation}\label{eq2}
\xi(k,r,\Delta)=\int_{-\infty}^{\infty}E_{k}(\omega)E^{\ast}_{r}(\omega+\Delta)d\omega=
\int_{-\infty}^{\infty}\left|E_{k}(\omega)\right|\cdot
\left|E_{r}(\omega+\Delta)\right|e^{i[\Phi_{k}(\omega)-\Phi_{r}(\omega+\Delta)]}d\omega.
\end{equation}
Neglecting the spectral envelope profiles of
$\left|E_{k,r}(\omega)\right|$, $\xi(k,r,\Delta)$ is formally
equivalent to $c(k,r,\tau)$, differing from it by $\Delta$ being
continuous variable representing detuning between spectral
components of $E_{k}(\omega)$ and $E_{r}(\omega+\Delta)$, while
$\tau$ is a discrete integer variable representing the relative
delay between the two codes.

Physically, $\xi(k,r,\Delta)$ interferes all the possible two-photon
pathways of one absorbed photon and one emitted photon which
spectral components are detuned by $\Delta$ within the Raman-type
process.
Suppose that variable $\omega$ is divided to small bins of size
$\Delta\omega$ and each bin is associated with specific value of $t$
- the index of the code component, so that the phase $\Phi(\omega)$
of the electric field $E(\omega)$ is constant across each bin.
For simplicity, we analize first the case where the spectral
envelope is extremely slow and can be approximated to be constant:
$\left|E_{k,r}(\omega)\right|\simeq1$,
Similar to $c(k,r,\tau)$, $\xi(k,r,\Delta)$ obtains its maximal
values when $\Delta=n\Delta\omega L$ if $k=r$, as the interference
between all the corresponding two-photon pathways is fully
constructive.
Here $n=0,1,2\ldots$ and $L$ is the period of the code.
For $\Delta\ne n\Delta\omega L$ or $k\ne r$ the interference between
the two-photon pathways is destructive and the value of
$\xi(k,r,\Delta)\ll \xi(k,k,n\Delta\omega L)$.
Overall, the profile of $\xi(k,r,\Delta)$ has peaks corresponding to
the cases of the constructive interference that appear with the
spacing associated with the period of $\Delta\omega L$.
If the interference is destructive, for the case of flat spectral
profiles of $E_{k,r}(\omega)$, the bounds on the value of
$\xi(k,r,\Delta)$ are obtained from the results of analysis of
$c(k,r,\tau)$ within the framework of Number Theory \cite{kumar_1,
kumar_2, kumar_3}.
For the proper choice of the codes family, these interference is
kept sufficiently low and enables clear discrimination between the
matching [$k=r$] and non-matching [$k\ne r$] codes.

Experimentally, multiple two-photon pathways are interfered within
the non-linear medium.
However, due to its Raman nature, the presented two-photon
photo-excitation is not applicable for the single-shot detection.
The latter is possible employing the four-wave mixing process within
the medium having third-order susceptibility $\chi^{(3)}$.
The interaction includes absorption of two-photons and emission of
one photon provided by the excitation electric fields of the lasers,
while the fourth emitted photon generated by the non-linear
polarization of the medium is detected:
\begin{equation}\label{eq3}
E_{sig}(\omega)\propto
\int_{-\infty}^{\infty}E_{pump}(\omega')d\omega'
\int_{-\infty}^{\infty}E_{k}(\omega'')E^{\ast}_{r}(\omega''+\omega'-\omega)d\omega''.
\end{equation}
To avoid the broadening of the interference pattern created by the
pulses $E_{k}(\omega)$ and $E_{r}(\omega)$ due to the presence of
the additional broadband pump laser pulse $E_{pump}(\omega)$, the
latter must be of narrow bandwidth.
Let us assume that
\begin{equation}\label{eq4}
E^{narrow}_{pump}(\omega)=\int_{-\infty}^{\infty}
E_{pump}(\omega')\delta(\omega_{0}-\omega')d\omega'.
\end{equation}
Then we obtain
\begin{eqnarray}\label{eq5}
E_{sig}(\omega_{0}+\Delta)&\propto&
E_{pump}(\omega_{0})\int_{-\infty}^{\infty}
E_{k}(\omega)E^{\ast}_{r}(\omega+\Delta)d\omega\\ \nonumber %
&=& E_{pump}(\omega_{0})\int_{-\infty}^{\infty}
\left|E_{k}(\omega)\right|\left|E^{\ast}_{r}(\omega+\Delta)\right|
e^{i[\Phi_{k}(\omega)-\Phi_{r}(\omega+\Delta)]}d\omega.
\end{eqnarray}
Therefore, we establish the recognition of matching codes and
measuring their period upon the measurement of the codes
cross-correlation function in the frequency domain, which is
equivalent to the measurement of the spectrum of $E_{sig}(\omega)$.
In the case of the finite bandwidth of $E_{pump}$ the
measured spectrum of $E_{sig}(\omega)$ represents the
cross-correlation function $\xi(k,r,\Delta)$ convoluted with
spectral profile of $E_{pump}$.

Our experimental setup utilizes ultrashort pulses of 65fs
(fwhm=15nm) produced by Ti:Sapphire oscillator at 800 nm that were
regeneratively amplified at 1KHz rate and splited into three beams.
Two beams were passed through the phase-shaping 4f optical set-up
incorporated 640-pixel liquid-crystal Spatial Light Modulator which
was used to assign different codes for each beam independently.
The corresponding experimental shaping resolution was
$\Delta\omega$=0.11 and 0.136 nm/pixel respectively.
The third beam was amplitude-shaped using similar 4f shaping set-up
to yield narrow [fwhm=0.9nm] spectrum.
The resulted three beams were focused at zero delay within the
BOXCARS geometrical configuration onto the 1mm-wide piece of fused
silica.
The resulted spatially isolated signal was collected and its
spectrum was measured by CCD camera coupled to the spectrometer with
resolution of 0.1nm.
It worth noting that alternatively to the presented BOXCARS
geometry, polarization gating or self-diffraction geometries can be
implemented within the same general formalism.

We have used the quaternary [$p$=4] codes which belong to the nested
families $S(0)$ and $S(1)$ and generated using corresponding linear
shift registers constructed upon the primitive polynomial.
The chosen period of the codes was $L=2^{l}-1$=7, where $l$=3 is the
order of the generating polynomial.
The size of the family $S(0)$ is $M=L$+2=9 with
$\left|C_{max}\right|$=$\sqrt{L+1}$+1$\simeq$3.8.
The size of the family $S(1)$ is $M=(L+2)(L+1)$=64.
Formally,  the corresponding value of $C_{max}$ for this case
$\left|C_{max}\right|$=$2\sqrt{L+1}$+1$\simeq$6.66, however, for the
specific case of $l$=3, occurrence of this value is zero and the
effective $\left|C_{max}\right|$=5.

\begin{figure} 
$$\includegraphics[width=8cm]{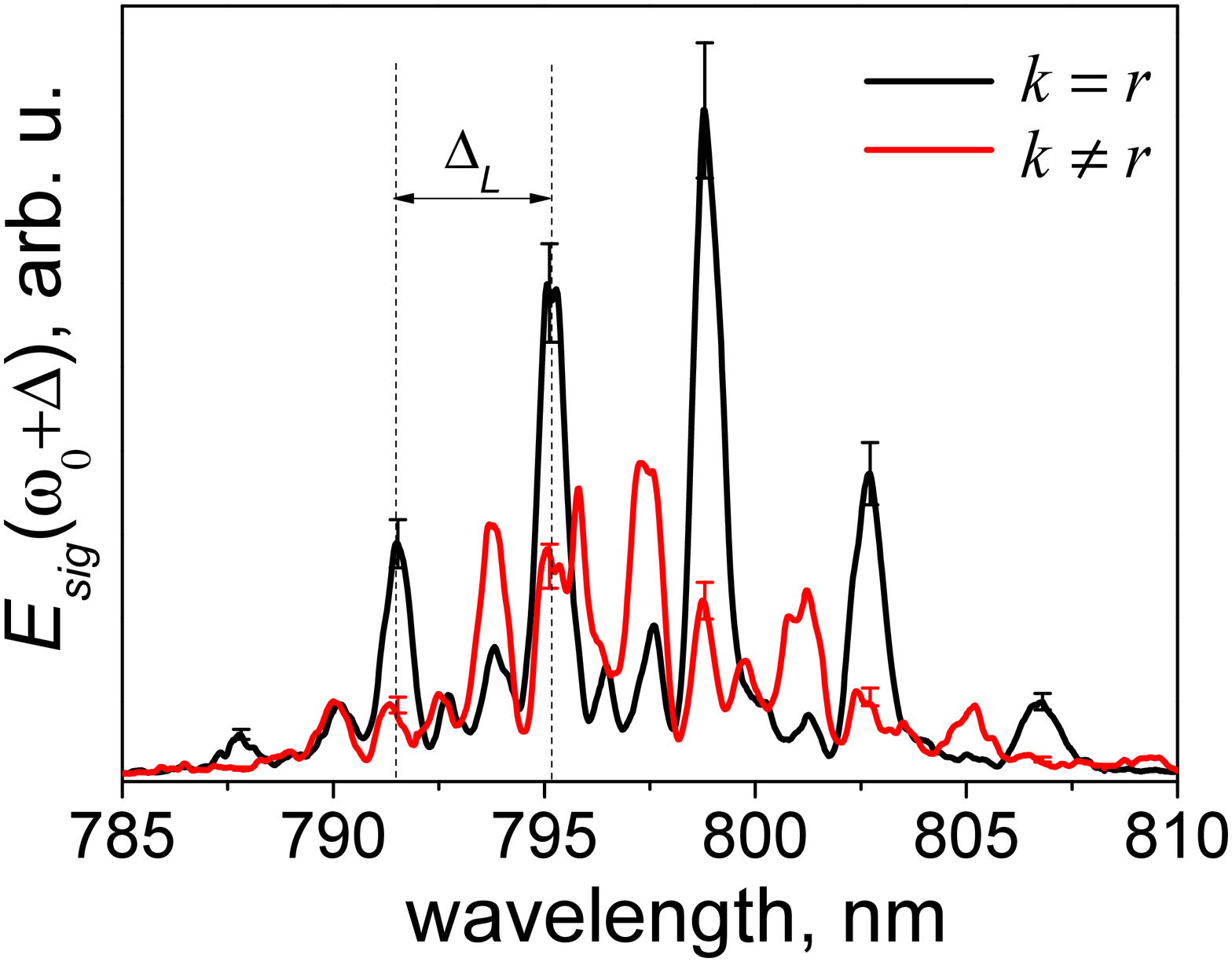}$$    
\caption{ %
Typical experimental spectrum of $E_{sig}(\omega)$ for the cases of identical [$k=r$, auto-correlation, black line] and different codes [$k \ne
r$, cross-correlation, red line].
The distance between the peaks for the case of $k=r$ corresponds to the period $\Delta_{L}$.
} \label{fig_1}
\end{figure}

\begin{figure} 
$$\includegraphics[width=8cm]{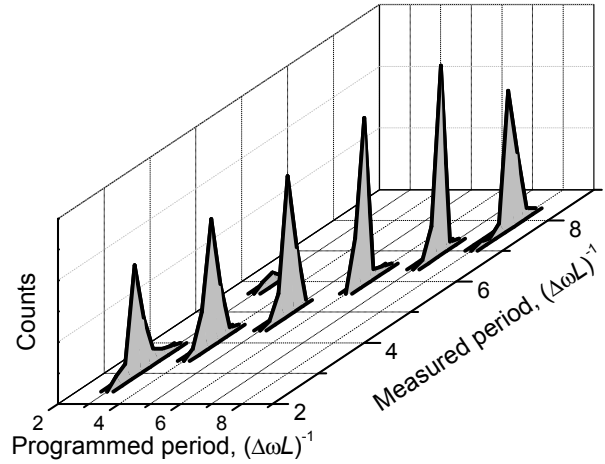}$$     
\caption{%
2D histogram of measured vs programmed code period $\Delta_{L}$ for 432 different codes belong to family $S(0)$ and $S(1)$.
The distribution along the right axis reflects the signal-to-noise experimental ratio, which results in the 98$\%$ true matching of the code
period.
%
} \label{fig_2}
\end{figure}

On Figure~\ref{fig_1} the typical spectrum of $E_{sig}(\omega)$ is
shown for the cases of $k=r$, and $k\ne r$.
The very pronounced peaks observed in the spectrum correspond to the
constructive interference between the photoinduced pathways
associated with the identical codes, where $k=r$.
The peaks are separated by $\Delta_{L}=n \Delta\omega L$, where $n$
- is the integer parameter.
This is in high contrast to the spectrum obtained for the case of
the destructive interference among the pathways for the case of
different codes, i.e. $k\ne r$.
For the peaks exceeding the certain threshold, the distances between
the peaks is used for the determination of the code period
corresponding to different values of $n$.
The corresponding results for the codes belong to the families
$S(0)$ and $S(1)$ are shown on Figure~\ref{fig_2} in the form of 2D
histogram, where $x$-axis corresponds to the programmed code period
and the $y$-axis -- to the retrieved period.
We have successively recognized periods of 98$\%$ out of total 432
different periodic codes.
As it is seen from Figure~\ref{fig_2} the origin of the 2$\%$ of the
false recognized codes comes from the shortest and longest periods
examined.
For the shortest period, the factor limiting the recognition power
is the actual width of the probe pulse, while its convolution smears
the spectral features and broadens the spectrum.
On the side of the longest periods, the total spectral width of
pulses $E_{k,r}(\omega)$ becomes important.
If the  pulse bandwidth is not broad enough, there is not enough
constructively interfering pathways provided and the contrast
between the cases of $k=r$ and $k\ne r$ is not sufficient for the
proper recognition.

\begin{figure} 
$$\includegraphics[width=13cm]{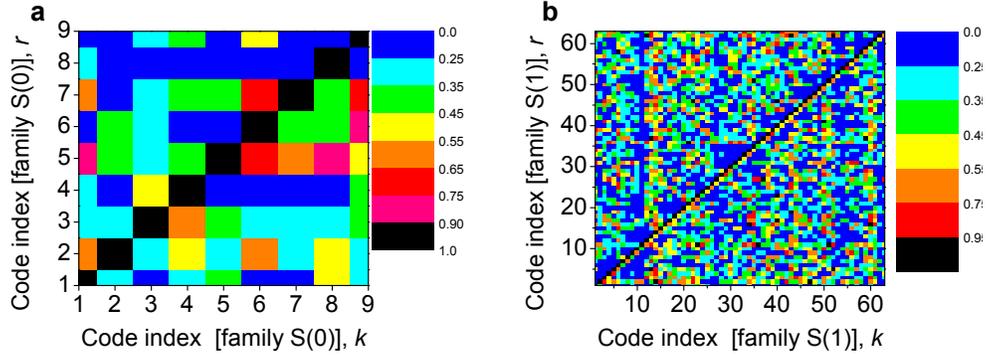}$$     
\caption{ %
Color-map plot representing the maximal cross-correlation values $\sum_{m}E_{sig}(\omega_{0}+m\Delta_{L})$, $m$=0,1,$\ldots$ for different
codes.
(a) - results for the family $S(0)$, $l$=3.
100$\%$ of the codes was efficiently matched based upon the cross-correlation measurements applying the threshold of 10$\%$.
(b) - family $S(1)$, $l$=3.
Applying the threshold of 95$\%$ we obtain that 0.6$\%$ out of whole set of measurements led to the false identification.
This result agrees with our numerical evaluation considering the experimental parameters.
} \label{fig_3}
\end{figure}

The codes with different periods are easily distinguished form the
measurement of $E_{sig}(\omega)$ due to the lack of the constructive
interference to the amplitude of the emitted signal.
For the experimental discrimination between different codes with the
same period it is enough to measure the signal at specific
frequencies $E_{sig}(\omega_{0}+m\Delta_{L})$, where $m=0,1,\ldots$.
We present the corresponding results for the code families $S(0)$
and $S(1)$ on the 2D color-map plot of Figure~\ref{fig_3} (a) and
(b) respectively.
Here the color represents the normalized value of
$\sum_{m}E_{sig}(\omega_{0}+m\Delta_{L})$ for the different pairs of
codes $k$ [$x$-axis] and $r$ [$y$-axis], so that the
auto-correlation cases appear on diagonal.
As the signal corresponding to $r=k$ is sufficiently higher than for
$r\ne k$, we obtain a very efficient recognition of the 100$\%$ of
codes belong to $S(0)$ with the signal threshold of 90$\%$ [see
panel (a)].
For the codes of the $S(1)$ family the value of $C_{max}$ is higher
then that of $S(0)$ and, hence, the threshold for the code
recognition must be also higher.
We obtain 0.6$\%$ of experimental false identification using
threshold of 95$\%$ in this case [see panel (b)] out of the whole
set of the cross-correlation measurements.
Our experimental results were verified with the numerical evaluation
of the spectrum of $E_{sig}(\omega)$ considering experimental
parameters as pulse-shaping resolution, bandwidth of the pump pulse,
etc.
The same fraction of the false identified signals was obtained.
We have also performed numerical simulation of the signal
considering the pulses with broader bandwidth [fwhm=25nm, pulse
duration 45fs], our experimental pulse-shaping resolution, and
bandwidth of the pump pulse of fwhm=0.8nm for the $S(1)$ family with
$l$=4.
Here $L$=15, $M$=255, while $\left|C_{max}\right|$=9.
For this case we obtain that 100$\%$ of codes can be recognized
applying the 90$\%$ threshold.

In conclusion, we have experimentally demonstrated the all-optical
single-shot frequency-domain cross-correlation measurements for the
efficient matching and simultaneously measuring the period of
low-correlating codes.
The presented technique has potential to be implemented in the
optical phase-shaped spread-spectrum code division multiple access
communication network.


\newpage

\end{document}